\documentclass[aps,preprint]{revtex4}%
\usepackage{amsfonts}
\usepackage{amsmath}
\usepackage{amssymb}
\usepackage{graphicx}%
\begin{document}
\preprint{ }
\title[accurate numerical spectral]{An accurate spectral method for solving the Schr\"{o}dinger equation.}
\author{G. H. Rawitscher}
\affiliation{Physics Department, University of Connecticut, Storrs, CT 06269-3046}
\author{I. Koltracht}
\affiliation{Mathematics Department, University of Connecticut, Storrs, CT 06269-3009}
\keywords{one two three}

\begin{abstract}
The solution of the Lippman-Schwinger (L-S) integral equation is equivalent to
the the solution of the Schr\"{o}dinger equation. A new numerical algorithm
for solving the L-S equation is described in simple terms, and its high
accuracy is confirmed for several physical situations. They are: the
scattering of an electron from a static hydrogen atom in the presence of
exchange, the scattering of two atoms at ultra low temperatures, and barrier
penetration in the presence of a resonance for a Morse potential. A key
ingredient of the method is to divide the radial range into partitions, and in
each partition expand the solution of the L-S equation into a set of Chebyshev
polynomials. The expansion is called "spectral" because it converges rapidly
to high accuracy. Properties of the Chebyshev expansion, such as rapid
convergence, are illustrated by means of a simple example.

\end{abstract}
\startpage{1}
\endpage{ }
\maketitle

\section{Introduction}

As stated in the textbook by Cummings, Laws, Redish and Cooney \cite{CLRC},
"Physics is a process of learning about the physical world by finding ways to
make sense of what we observe and measure. As the inspiring teacher Richard
Feynman wrote, \cite{Feynman} "Progress in all of the natural sciences depends
on this interaction between experiment and theory"."

An important tool required for carrying out this interaction is the solution
of equations provided by a particular theory, in order to be able to compare
its predictions with experiment. As the equations become more and more
involved, such as in global climate study, in the construction of
pharmaceutical drugs, in the analysis of large organic chains that exist in
live cells, in the understanding of superconductivity, in the tracing of the
earth's interior by means of seismic waves, in the construction of devices
that transmit digital information, in the study of atomic, nuclear and
particle theory (particularly in lattice gauge theory), etc., the resort to
numerical computational methods becomes increasingly more necessary.

The purpose of this paper is to point out special physical situations that
require very accurate numerical algorithms, and to describe one such algorithm
that has been recently developed. These special cases require either the
evaluation of the solution of a wave equation out to large distances, or
require high accuracy even for small distances, or both. Examples are the
collision between atoms at extremely low temperatures. The understanding of
such collisions is important for astro-physical applications, for the
description of the state of atoms or molecules called Bose-Einstein
condensates, and for the understanding of superfluidity in liquids formed out
of weakly interacting atoms, such as the atoms of Helium. Helium is a "noble
gas", i.e., its atoms interact mainly repulsively at short distances, yet, at
intermediate distances (between 5 and 200 atomic units of distance) there is a
small attractive valley in the potential energy curve (of a depth less than
$3.5\times10^{-3}$ atomic units of energy) within which a bound state can
form. That weak attraction is in turn important for the molecular binding of a
system of three or more helium atoms \cite{He3-1}, \cite{He3-2}. The quantum
mechanical wave function for the di-atom, in view of the weak binding energy
of $4.4\times10^{-9}$ atomic units of energy \cite{He-He-1}, extends to such
large distances that accurate numerical values out to 2000 atomic units are required.

For the case of the radial, one-dimensional, Schr\"{o}dinger equation
\begin{equation}
\left(  d^{2}/dr^{2}+k^{2}\right)  \psi=V\,\psi, \label{SCHR}%
\end{equation}
where $k$ is the wave number in units of inverse length and $V(r)$ is the
potential in units of inverse length squared which contains the $L(L+1)/r^{2}$
singularity, the most suitable equivalent integral equation for the S-IEM
method is the Lippman-Schwinger equation
\begin{equation}
\psi(r)=\sin(kr)+\int_{0}^{T}\mathcal{G}_{0}(r,r^{\prime})\,V(r^{\prime
})\,\psi(r^{\prime})\,dr^{\prime}, \label{LS}%
\end{equation}
where $\mathcal{G}_{0}$ is the undistorted Green's function. In configuration
space $\mathcal{G}_{0}$ has the well known semi-separable form $\mathcal{G}%
_{0}=-(1/k)\,\sin(kr_{<})\,\cos(kr_{>}).$ (for negative energies one would
have $-(1/\kappa)\,\sinh(\kappa r_{<})\,\exp(-\kappa r_{>}))$. By introducing
the integral operator $\mathcal{K}_{T},$ so that when applied on a function
$\psi(r)$ the result is%
\begin{equation}
\mathcal{K}_{T}\psi(r)\equiv-\frac{1}{k}\cos(kr)\int_{0}^{r}dr^{\prime}%
\ \sin(kr^{\prime})\ V(r^{\prime})\psi(r^{\prime})-\frac{1}{k}\sin(kr)\int
_{r}^{T}dr^{\prime}\ \cos(kr^{\prime})\ V(r^{\prime})\psi(r^{\prime}),
\label{IOT}%
\end{equation}
then Eq. (\ref{LS}) can be written as
\begin{equation}
\psi(r)=\sin(kr)+\mathcal{K}_{T}\,\psi(r), \label{LSK}%
\end{equation}
where $\mathcal{K}_{T}\,\psi$ means that $\psi(r^{\prime})$ is included in the
integrands contained in Eq. (\ref{IOT}). This form of Eq. (\ref{LSK}) leads to
the boundary condition that $\psi(0)=0$, and since it assumes that for $r\geq
T$ the potential $V(r)=0,$ it leads to the asymptotic behavior $\psi
(r)=\sin(kr)+B\ \cos(kr)$, where $B$ is a constant determined from the
solution of Eq. (\ref{LS}). If $V(r)\neq0$ for $r\geq T$, then matching at
$r=T$ to the corresponding long range functions (Bessel or Coulomb, for
example) is required, as is explained in Ref. (\cite{IEM1}, \cite{IEM2}).

A new method for solving the Lippman-Schwinger integral equation (\ref{LS}),
associated with the differential Schr\"{o}dinger equation (\ref{SCHR}), has
been developed recently \cite{IEM1} as an extension of a method due to
Greengard and Rokhlin \cite{FR}. This method, to be called IEM\ (for integral
equation method) has an accuracy which, for the same number of mesh-points, is
far superior to the accuracy provided by finite difference methods for solving
either an integral or a differential equation. One of the intended
applications \cite{fb17} is the solution of the Faddeev equations for a
three-body system in configuration space, since it requires the calculation of
wave functions out to large distances. It is the purpose of this paper to
describe the application of this method for positive energy, two-body
scattering cases, and compare it with several other methods. The application
of this method to finding bound-state negative energies is being developed,
with the intention of obtaining the He-He bound state described above. The
basic idea of the IEM is to divide the radial interval into partitions, obtain
two special solutions of the restricted Lippman-Schwinger equation in each
partition, called $Y(r)$ and $Z(r),$ by expanding these solutions into a set
of Chebyshev polynomials, and calculating the coefficients of the expansion in
each partition. That expansion is "spectral", i.e., it converges rapidly once
the number of terms exceeds a certain value, and the error of truncating the
expansion beyond that value is known, as is further explained below. Once the
functions $Y$ and $Z$ are obtained in each partition, then the global function
$\psi$ in that partition is expressed as a linear combination of the $Y$ and
$Z$ . The coefficients of that combination are subsequently calculated by
solving a matrix equation, which is sparse, as will be explained. Spectral
expansions to solve integral equations, albeit using a rather different
set-up, in particular not using Green's functions or partitions, has also
recently been developed by B. Mihaila \cite{BM}.

Even though it is known that the errors which arise in the\ numerical solution
of an integral equation are smaller than the errors in the solution of an
equivalent differential equation, it is customary to solve the latter. The
reason is that the algorithms for solving a differential equation by means of
finite difference methods (such as Numerov of Runge-Kutta) are simple and do
not require extensive storage space. By contrast, the discretization of an
integral equation usually leads to large non-sparse matrices, and hence
requires large investments of computer time and storage space. Therefore the
gain in accuracy of the integral equation formulation is normally offset by a
manifold increase in computational time. Our method circumvents this problem,
as is described below. Before applications to physical cases are described, it
is instructive to understand the basic accuracy properties of the spectral
expansion method, as well as the basic ingredients of the IEM.

\section{Spectral Expansion}

The main feature of a spectral expansion, namely its rapid convergence, will
now be demonstrated by means of a simple example even though extensive
discussions exist in the literature \cite{GOTT}. For the spectral expansion
functions we will use Chebyshev polynomials only, although other orthogonal
polynomials, such as Legendre, are also often used. We use Chebyshev
polynomials because they are particularly well suited for obtaining the
antiderivaties that appear in Eq. (\ref{IOT}).

Spectral accuracy is described as follows: If a function $f(x),~-1\leq x\leq1$
is expanded in terms of Chebyshev polynomials $T_{j}(x)$,
\begin{equation}
f(x)=\frac{a_{0}}{2}+\sum_{j=1}^{\infty}a_{j}T_{j}(x) \label{EXP}%
\end{equation}
then the error in truncating the expansion after $n$ terms is proportional to
$(n+1)^{-p},$ where $\ p$ is the number of continuous derivatives which the
function $f$ \ has in the in the interval $-1<x<1.$ Furthermore, this
truncation error is also proportional to the $(n+1)^{\prime}$th coefficient of
the expansion, which means that, after a certain number of terms, the
coefficients $a_{j}$ decrease rapidly with $j$ according to the same law
$j^{-p}$. In particular, if $f(x)$ is infinitely differentiable, then the
coefficients $a_{i}$ converge to zero asymptotically faster than any fixed
power of $(1/j).$ Hence the term \textquotedblleft spectral
convergence\textquotedblright\ is also referred to as \textquotedblleft
superalgebraic convergence\textquotedblright.

These properties will now be illustrated by expanding the function
$f(x)=\exp(x)$ into Chebyshev polynomials. The coefficients $a_{j}$ in Eq.
(\ref{EXP}) are given by%
\begin{equation}
a_{j}=\frac{2}{\pi}\int_{-1}^{1}e^{x}T_{j}(x)(1-x^{2})^{-1/2}dx=\frac{2}{\pi
}\int_{0}^{\pi}~e^{\cos\theta}\cos(j\theta)d\theta, \label{EXPC}%
\end{equation}
which follows from the orthogonality relation
\begin{align}
\int_{-1}^{1}T_{k}(x)\ T_{j}(x)\ (1-x^{2})^{1/2}\ dx  &  =0\text{ ~~~~if
}j\neq k\nonumber\\
&  =\pi/2\text{ if~ }j=k\neq0\nonumber\\
&  =\pi~~~~~\text{if~\ }j=k=0. \label{ORTN}%
\end{align}
The integral in Eq. (\ref{EXPC}) can be calculated analytically. In view of
Eq. (6.9.19) in Ref. \cite{AS} the result is $a_{j}=2I_{j}(1)$, where
$I_{j}(z)$\ is a modified Bessel function of order $j$. Using the asymptotic
expansion for large orders of a Bessel function, Eq. (9.3.1) of Ref.
\cite{AS}, an approximation to $a_{j}$ for large values of the index $j$ \ is
\begin{equation}
a_{j}\simeq\frac{2}{\sqrt{2\pi j}}\left(  \frac{e}{2j}\right)  ^{j}%
;\ \ \ \ \ \ \ \ j\rightarrow\infty\label{EXPA}%
\end{equation}
Equation (\ref{EXPA}) shows that the value of $a_{j}$ decreases with $j$
faster than any fixed power of $j,$ as is also demonstrated in the Table
\ref{TAB1}.
\begin{table}[tbp] \centering
\begin{tabular}
[c]{||c||c||c||c||c|}\hline
& $a_{2}$ & $a_{4}$ & $a_{6}$ & $a_{8}$\\\hline\hline
$\ Eq.(\ref{EXPC})$ & $2.715E-1$ & $5.474E-3$ & $4.450E-5$ &
\multicolumn{1}{||c||}{$1.992E-7$}\\
$\ Eq.(\ref{EXPA})$ & $2.60E-1$ & $5.32E-3$ & $4.40E-5$ &
\multicolumn{1}{||c||}{$1.958E-7$}\\\hline
\end{tabular}
\caption{Chebyshev expansion coefficients a(k) of f(x)=exp(x)\label{TAB1}}
\end{table}%
. The first row lists the values of $a_{j}$ for $j=2,4,6,8$ as calculated from
Eq. (\ref{EXPC}), (the results for the odd values of $j$ are not shown) and
the second row gives the values obtained from the asymptotic approximation
(\ref{EXPA}) The table shows that the coefficients decrease rapidly with the
order $j.$ Will the truncation error also decrease rapidly?

\begin{figure}
[ptb]
\begin{center}
\includegraphics[
natheight=2.885000in,
natwidth=3.837300in,
height=3.0178in,
width=4.005in
]%
{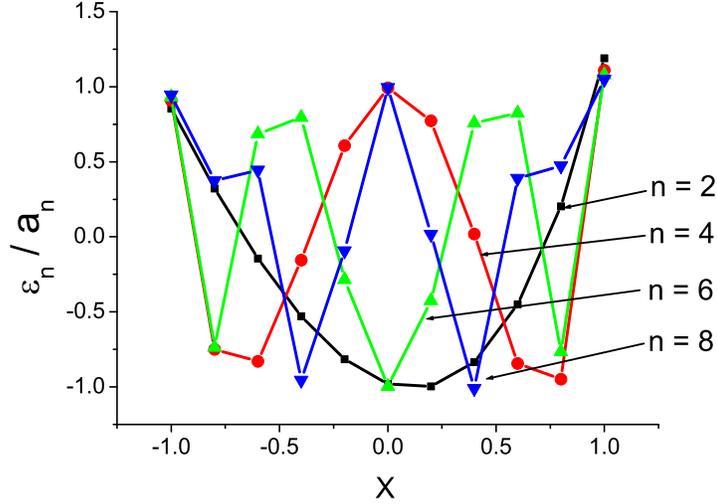}%
\caption{Truncation errors in the expansion of $f(x)=exp(x)$ into Chebyshev
polynomials, divided by the first expansion coefficient not included in the
sum.}%
\label{FIGC}%
\end{center}
\end{figure}

The truncation error in the expansion is defined as $\epsilon_{n}%
(x)=f(x)-f_{n}(x)$ where $f_{n}(x)$ denotes the sum in Eq. (\ref{EXP}) that is
taken from $j=1$ to $j_{\max}=n-1.$ A useful property of spectral expansions
is that this error decreases with $n$ proportionally to $a_{n},$ the first
expansion coefficient not included in the sum. This is demonstrated in Fig.
\ref{FIGC}, which shows the ratio $\epsilon_{n}(x)/a_{n}$, for $n=2,4,6$ and
$8.$ The figure shows that the curves are approximately contained between
$\pm1$, i.e., the truncation error is of the same magnitude as $a_{n}$
independently of the value of $x$. Hence the truncation error does not show a
Gibbs phenomenon at the end points, as would be the case for an expansion into
a Fourier Series.\ %

The above mentioned relation between the truncation error and the value of the
Chebyshev coefficient provides a convenient method for finding the appropriate
size of each partition, compatible with the overall prescribed error. Clenshaw
and Curtis \cite{CC}, who originated this spectral integration technique,
recommend using the average size of the three last consecutive coefficients as
an accuracy criterion.

Once the coefficients $a_{i}$ of the expansion (\ref{EXP}) are known for
$j=0,1,..N$, then one has a semi-analytical approximation to the function
$f(x),$ given by the truncated form of Eq. (\ref{EXP})
\begin{equation}
f_{N}(x)=\frac{a_{0}}{2}+\sum_{j=1}^{N}a_{j}T_{j}(x), \label{EXPN}%
\end{equation}
that enables one to evaluate $f_{N}$ at any point $x$ in the interval
$[-1,+1]$ without the need to carry out interpolations. A method for obtaining
the coefficients $a_{j}$ that does not require to evaluate the integrals in
Eq. (\ref{EXPC}) is described in Ref. \cite{CC}. It consists in considering
the $N+1$ zeros $\xi_{a}$ of $T_{N+1}$ for $\alpha=0,1,..N,$ evaluating the
expansion (\ref{EXPN}) at $x=\xi_{a}$ for $\alpha=0,1,..N$ and thus obtaining
a set of $N+1$ linear equations for the coefficients $a_{j}$. The matrix
involved that relates the column vector of the $f(\xi_{a})$ to the vector of
the $a_{j}$ has elements formed from the values $T_{j}(\xi_{a}),$ with
$j,\alpha=0,1,..N.$. Details can be found in Ref. \cite{IEM1} and in
textbooks. This is the method used to construct Tables \ref{TAB2}-\ref{TAB4}.

The Chebyshev expansion is particularly suited to obtain the integral
$\int_{-1}^{x}f_{N}(x^{\prime})~dx^{\prime}$ of the function $f_{N}$ without
significant loss of accuracy$.$ An expansion of this antiderivative function
in terms of Chebyshev polynomials
\begin{equation}
F_{N}(x)=\int_{-1}^{x}f_{N}(x^{\prime})~dx^{\prime}=\sum_{j=0}^{N+1}b_{j}%
T_{j}(x). \label{EXPI}%
\end{equation}
has the property that the coefficients $b_{j}$ can be easily obtained in terms
of the coefficients $a_{j}$, by means of a matrix usually denoted as $S_{L}$,
as is described in textbooks as well as in Ref. \cite{IEM1}. The basic reason
is that the integral from $-1$ to $x$ of a particular $T_{j}$ is given by a
linear combination of $T_{i}(x)$ with $i\ \leq\ j+1$. For example, $\int
_{-1}^{x}\ T_{2}(x^{\prime})\ dx^{\prime}=[T_{3}(x)-3T_{1}(x)-2T_{0}(x)]/6$,
and $\int_{-1}^{x}\ T_{3}(x^{\prime})\ dx^{\prime}=[T_{4}(x)-2T_{2}%
(x)+T_{0}(x)]/8$. The sum in Eq. (\ref{EXPI}) should rigorously go to the
upper limit $N+1.$ However, in numerical calculations the $(N+1)$'th term is
generally ignored. A similar matrix, called $S_{R},$ exists in order to obtain
a Chebyshev expansion of $\int_{x}^{1}f_{N}(x^{\prime})~dx^{\prime}$ A
numerical verification that the accuracy of the antiderivative is of the same
order of magnitude as the accuracy of the expansion of the function $f_{N}$,
again for $f(x)=\exp(x)$, is shown in the second and third columns of Table
\ref{TAB2}.
\begin{table}[tbp] \centering
\begin{tabular}
[c]{|c||c|c|c|c|}\hline
$x$ & $F_{N}-e^{x}+e^{-1}$ & $f_{N}-e^{x}$ & $f_{N}^{(1)}-e^{x}$ &
$f_{N}^{(2)}-e^{x}$\\\hline\cline{2-5}%
$-0.8$ & $.25(-11)$ & $-.51(-9)$ & $.12(-8)$ & $.14(-6)$\\\hline
$-0.6$ & $.11(-8)$ & $.51(-9)$ & $.10(-8)$ & $-.81(-7)$\\\hline
$-0.4$ & $.29(-9)$ & $-.30(-9)$ & $-.48(-8)$ & $.37(-7)$\\\hline
$-0.2$ & $.27(-9)$ & $-.23(-9)$ & $.49(-8)$ & $.24(-7)$\\\hline
$0.0$ & $.11(-8)$ & $-.55(-9)$ & $.50(-10)$ & $-.55(-7)$\\\hline
$0.2$ & $.37(-9)$ & $-.24(-9)$ & $-.52(-8)$ & $.23(-7)$\\\hline
$0.4$ & $.20(-9)$ & $-.32(-9)$ & $.51(-8)$ & $.41(-7)$\\\hline
$0.6$ & $.11(-8)$ & $.57(-9)$ & $-.10(-8)$ & $-.91(-7)$\\\hline
$0.8$ & $.21(-10)$ & $-.58(-9)$ & $-.15(-8)$ & $.16(-6)$\\\hline
\end{tabular}
\caption{ Coefficients $a_{j}$ and $a_{j}\times j^{2}$
for the expansion of $exp(x)$ for $N=9$  \label{TAB2}}%
\end{table}%
\medskip

The derivatives with respect to $x$ of $f_{N}$ \ can also be obtained via
Chebyshev expansions, but in order to maintain a prescribed accuracy, the
truncation value $N$ has to be inreased accordingly. Call $f_{N}^{(1)}%
=df_{N}/dx,\ f_{N}^{(2)}=d^{2}f_{N}/dx^{2},\ etc.$ One of two methods consists
in taking the derivatives of the Chebyshev polynomials term by term in Eq.
(\ref{EXPN})
\begin{equation}
f_{N}^{(n)}(x)=\sum_{j=1}^{N}a_{j}T_{j}^{(n)}(x),~~~n=1,2,... \label{EXPD}%
\end{equation}
The expressions for $T_{j}^{(n)}(x)$ can be given analytically, and hence
$f_{N}^{(n)}$ can be evaluated numerically at any point $x$ in $[-1,+1].$ By
taking a derivative of a polynomial of order $j,$ the result is a polynomial
of order $j-1,$ whose magnitude is of order $j$ times the original polynomial.
For example, $d^{2}T_{j}(x)/dx^{2}=[xdT_{j}/dx-j^{2}T_{j}]/(1-x^{2}).$That
leads one to expect that the errors in Table \ref{TAB2} for a derivative of
order $n$ are related to the coefficient of the next to the last Chebyshev
polynomial, ($T_{N+1}$) times $(N+1)^{n}.$ Table \ref{TAB3} lists coefficients
$a_{i}$ and $a_{j}\times j^{2}$
\begin{table}[tbp] \centering
\begin{tabular}
[c]{|c|c|c|c|c|c|}\hline
$j$ & $7$ & $8$ & $9$ & $10$ & $11$\\\hline
$a_{j}$ & $.32(-5)$ & $.20(-6)$ & $.11(-7)$ & $.55(-9)$ & $.25(-10)$\\\hline
$a_{j}\times j^{2}$ & $.16(-3)$ & $.13(-4)$ & $.88(-6)$ & $.55(-7)$ &
$.30(-8)$\\\hline
\end{tabular}
\caption{ Coefficients $a_{j}$ and $a_{j}\times j^{2}$ for
the expansion of $exp(x)$ \label{TAB3}}%
\end{table}
and by comparing Tables \ref{TAB2} and \ref{TAB3} one sees that this
expectation is borne out.

A second method consists in writing a Chebyshev expansion for $df/dx$%
\begin{equation}
df_{N}/dx=\frac{c_{0}}{2}+\sum_{j=1}^{N-1}c_{j}T_{j}(x), \label{EXPD1C}%
\end{equation}
and by noting that the expansion coefficients $c_{j}$ are related to the
coefficients $a_{j}$ in Eq. (\ref{EXPN}) as\ follows: $c_{N-1}=2Na_{N},$
$c_{N-2}=2(N-1)a_{n-1},$ and for $j\leq N-2,$ $c_{j-1}=c_{j+1}+2ja_{j}.$ The
error in $df_{N}/dx$ is approximately equal to the magnitude of $c_{N},$ that
in turn permits one to determine the value of $N$ from the relation
$c_{N}=2(N+1)a_{N+1}$

In the numerical example given in this section the upper value $N$ of the sums
in the Chebyshev expansions was taken as $N=9$. However, in the numerical
solution of the integral equation, as described in the next section, $N=15$.
This leads to accuracies of the order of $10^{-14}$, as is discussed in the
realistic numerical examples described below. In order to demonstrate the
rapid gain in accuracy for a small increase in the value of $N$, we show
errors similar to those displayed in Table \ref{TAB2}, for $N=13.$
\begin{table}[tbp] \centering
\begin{tabular}
[c]{|c|c|c|c|c|}\hline
$x$ & $F_{N}-e^{x}+e^{-1}$ & $f_{N}-e^{x}$ & $f_{N}^{(1)}-e^{x}$ &
$f_{N}^{(2)}-e^{x}$\\\hline
$-0.8$ & $.29(-14)$ & $.72(-15)$ & $.24(-14)$ & $-.44(-12)$\\\hline
$-0.6$ & $.36(-15)$ & $.33(-15)$ & $.21(-13)$ & $.12(-12)$\\\hline
$-0.4$ & $.31(-14)$ & $.44(-15)$ & $-.32(-13)$ & $-.15(-12)$\\\hline
$-0.2$ & $-.56(-16)$ & $-.22(-14)$ & $.29(-13)$ & $.37(-12)$\\\hline
$0.0$ & $.33(-14)$ & $.48(-14)$ & $-.11(-13)$ & $-.57(-12)$\\\hline
$0.2$ & $..11(-15)$ & $-.40(-14)$ & $-.21(-13)$ & $.57(-12)$\\\hline
$0.4$ & $.24(-14)$ & $.20(-14)$ & $.44(-13)$ & $-.37(-12)$\\\hline
$0.6$ & $.67(-15)$ & $0$ & $-.42(-13)$ & $.25(-12)$\\\hline
$0.8$ & $.29(-14)$ & $.88(-15)$ & $.19(-13)$ & $-.69(-12)$\\\hline
\end{tabular}
\caption{Same as Table \ref{TAB2}. for $N=13$ \label{TAB4}}%
\end{table}%
The accuracy increases approximately by four or five orders of magnitude as
$N$ is increased from $9$ to $13$.\medskip

Once the coefficients of a Chebyshev expansion (\ref{EXPN})of a function
$f_{N}(x)$ are obtained, the Fourier components $\int_{-1}^{+1}f(x)\sin(ax)dx$
and $\int_{-1}^{+1}f(x)\cos(ax)dx$ of that function can also be obtained, as
follows. If the coefficients $d_{k}$ of the expansion of the function
\begin{equation}
f(x)\ \sin(ax)=\sum_{k=0}^{M}\ d_{k}T_{k}(x) \label{1}%
\end{equation}
are known, then the integrals $\int_{-1}^{+1}\ f(x)\ \sin(ax)\ dx$ can be
easily obtained by applying the matrix $S_{L}$ described above upon the row
vector of the coefficients $d_{k}$, and remembering that $T_{k}(1)=1.$ In
order to obtain the coefficients $d_{k}$ one requires the integral%
\begin{equation}
\int_{-1}^{+1}T_{k}(x)\ f(x)\ \frac{\sin(ax)}{\sqrt{1-x^{2}}}dx=\sum_{j}%
a_{j}\int_{-1}^{+1}T_{k}(x)\frac{\sin(ax)}{\sqrt{1-x^{2}}}T_{j}(x)\ dx,
\label{2}%
\end{equation}
in view of Eqs. (\ref{ORTN}). By using the relation
\begin{equation}
2T_{k}(x)T_{j}(x)=T_{k+j}(x)+T_{|k-j|}(x) \label{3}%
\end{equation}
the integrals on the right hand side of Eq. (\ref{2}) can be carried out
analytically in terms of Bessel $J$ functions by using the expression
\cite{GRADR}%
\begin{equation}
\int_{-1}^{1}\ T_{2n+1}(x)\frac{\sin(ax)}{\sqrt{1-x^{2}}}dx=(-1)^{n}\pi
J_{2n+1}(a). \label{4}%
\end{equation}
For Chebyshev polynomials of even order the above integrals vanish.
Similarily, one can obtain the coefficients of the Chebyshev expansion of
$f_{N}(x)\cos(ax)$ by making use of \cite{GRADR}
\begin{equation}
\int_{-1}^{1}\ T_{2n}(x)\frac{\cos(ax)}{\sqrt{1-x^{2}}}dx=(-1)^{n}\pi
J_{2n}(a) \label{5}%
\end{equation}
In this manner the loss of accuracy in the integrals above that takes place
for large values of $a$ can be avoided.

Finally, we remark that the Chebyshev expansions can be used on any interval
$[a,b]$ by means of the linear transformation%
\begin{equation}
x=\frac{2}{b-a}r-\frac{b+a}{b-a} \label{6}%
\end{equation}
that maps $r\in\lbrack a,b]$ into $x\in\lbrack-1,1].$

\bigskip

\section{The Integral Equation method}

Our method for solving the Lippman-Schwinger equation (\ref{LS}) is described
below for the case of one channel and positive energy. The boundary
conditions, and hence the choice of the Green's function, is appropriate for a
scattering situation. Beyond a large radial distance called $T$ the potential
other than the centripetal or Coulomb potentials is set to zero. The radial
interval $[0,T]$ is partitioned into subintervals $i$, with $i=1,2,...M$. The
lower and upper boundaries of interval $i$ are $b_{i-1}$ and $b_{i}$,
respectively, with $b_{M}=T.$ In each partition the integral operator
$\mathcal{K}_{i}$ is defined%
\begin{equation}
\mathcal{K}_{i}=-\frac{1}{k}\cos(kr)\int_{b_{i-1}}^{r}dr^{\prime}%
\ \sin(kr^{\prime})\ V(r^{\prime})-\frac{1}{k}\sin(kr)\int_{r}^{b_{i}%
}dr^{\prime}\ \cos(kr^{\prime})\ V(r^{\prime}),~~b_{i-1}\leq r\leq b_{i}.
\label{IOi}%
\end{equation}
This operator is similar to $\mathcal{K}_{T}$ defined in Eq. (\ref{IOT}), with
the exception that the upper and lower limits of the integration are $b_{i-1}$
and $b_{i}$. Two independent local solutions $Y_{i}(r)$ and $Z_{i}(r)$\ in
partition $i$ are obtained by solving the integral equation locally, driven by
two different functions $\sin(kr)$ and $\cos(kr),$%
\begin{align}
(1-\mathcal{K}_{i}\mathcal{)}Y_{i}\  &  \mathcal{=}\sin(kr);~~~\ b_{i-1}\leq
r\leq b_{i}\nonumber\\
(1-\mathcal{K}_{i}\mathcal{)}Z_{i}\  &  \mathcal{=}\cos(kr);~~~\ b_{i-1}\leq
r\leq b_{i}. \label{YZ}%
\end{align}
It is important to note that boundary conditions are not needed to make the
solutions of Eqs. (\ref{YZ}) unique, unless the operator $(1-\mathcal{K}_{i})$
has zero eigenvalues. This situation is of course different from the solutions
of the differential equation (\ref{SCHR}), since the functions $\sin(kr)$ and
$\cos(kr)$ are eigenvectors of the operator $\left(  d^{2}/dr^{2}%
+k^{2}\right)  $ corresponding to zero eigenvalue. If accidentally the
operator $(1-\mathcal{K}_{i})$ has a zero eigenvalue in a particular
partition, then by decreasing the size of the partition the zero eigenvalue
should disappear because the "size" of $\mathcal{K}_{i}$ decreases
correspondingly. Another advantage of the integral equation method over the
differential equation method is that the operator $\mathcal{K}_{i}$ is
compact, while the operator $\left(  d^{2}/dr^{2}+k^{2}\right)  $ is not. A
compact operator can be approximated to ever increasing accuracy by a
separable expansion of basis vectors, and hence a numerical representation (or
discretization) of the operator is numerically stable.

The values of the functions $Y$ and $Z$ \ and their derivatives at the
boundary points of the partition $i$ can be obtained from Eqs. (\ref{YZ}) by
inserting into Eq. (\ref{IOi}) for $r$ the value $b_{i-1}$ or $b_{i},$
respectively. By defining the dimensionless quantities
\begin{align}
(GY)_{i}  &  =\frac{1}{k}\int_{b_{i-1}}^{b_{i}}\cos(kr)V(r)Y_{i}%
(r)dr\ ;~~~(FY)_{i}=\frac{1}{k}\int_{b_{i-1}}^{b_{i}}\sin(kr)V(r)Y_{i}%
(r)dr\nonumber\\
(GZ)_{i}  &  =\frac{1}{k}\int_{b_{i-1}}^{b_{i}}\cos(kr)V(r)Z_{i}%
(r)dr\ ;~~~(FZ)_{i}=\frac{1}{k}\int_{b_{i-1}}^{b_{i}}\sin(kr)V(r)Z_{i}(r)dr
\label{OYZ}%
\end{align}
one obtains%
\begin{align}
Y_{i}(b_{i-1})  &  =\sin(kb_{i-1})[1-(GY)_{i}]~~\nonumber\\
Y_{i}^{\prime}(b_{i-1})  &  =k\cos(kb_{i-1})[1-(GY)_{i}]\nonumber\\
Z_{i}(b_{i-1})  &  =\cos(kb_{i-1})-\sin(kb_{i-1})(GZ)_{i}]~~\nonumber\\
Z_{i}^{\prime}(b_{i-1})  &  =-k[\sin(kb_{i-1})+\cos(kb_{i-1})(GZ)_{i}]
\label{Bi-1}%
\end{align}

and
\begin{align}
Y_{i}(b_{i})  &  =\sin(kb_{i})-\cos(kb_{i})(FY)_{i}\ ~~\nonumber\\
Y_{i}^{\prime}(b_{i})  &  =k[\cos(kb_{i})+\sin(kb_{i})(FY)_{i}]\nonumber\\
Z_{i}(b_{i})  &  =\cos(kb_{i})[1-(FZ)_{i}]\ ~~\nonumber\\
Z_{i}^{\prime}(b_{i-1})  &  =-k\sin(kb_{i-1})[1-(FZ)_{i}]. \label{Bi}%
\end{align}
In the above, a prime indicates a derivative with respect to $r.$ Since the
functions $Y$ and $Z$ obey the Schr\"{o}dinger equation (\ref{SCHR}), the
wronskian of these functions, $W(Y,Z)=Y^{\prime}Z-YZ^{\prime},$ is independent
of the point $r$ within the interval $i$ if $V$ is a local potential. Using
the Eqs. (\ref{Bi-1}) and (\ref{Bi}) one can express the wronskian at
$r=b_{i-1}$ and $r=b_{i},$ respectively, in terms of the overlap integrals
defined in Eq. (\ref{OYZ}). One obtains%
\begin{align}
W(Y,Z)_{b_{i-1}}  &  =k[1-(GY)_{i}]\nonumber\\
W(Y,Z)_{b_{i}}  &  =k[1-(FZ)_{i}], \label{WRONSK}%
\end{align}
which implies in particular that
\begin{equation}
(GY)_{i}=(FZ)_{i}. \label{IDENT}%
\end{equation}
This result also shows that if $(GY)$ becomes close to unity in a particular
partition, then the functions $Y$ \ and $Z$ will no longer be significantly
linearly independent of each other, and the IEM\ method becomes unreliable in
this partition. The remedy is to decrease the length of the partition, since
the value of $(GY)$ will then also decrease.

The solution of Eqs. (\ref{YZ}) in each interval $i$ is accomplished by
expanding these functions in terms of Chebyshev Polynomials, and solving the
matrix equations for the corresponding coefficients. The procedure is well
described in Ref. \cite{IEM1}, and will not be repeated here. However, a few
remarks are in order: 1. The coefficients of the expansion of the functions
$Y_{i}(r)$ and $Z_{i}(r)$\ in terms of the Chebyshev polynomials are obtained
with high spectral accuracy by using Chebyshev collocation points in each
partition, together with the Curtis-Clenshaw quadrature \cite{CC}. 2. The Eqs.
(\ref{YZ}) are not the inverse of the Schr\"{o}dinger Eq., otherwise there
would be no gain in accuracy in using the integral equation. 3. The inverse of
the operator $(1-\mathcal{K}_{i}\mathcal{)}$ always exists if the partition
$i$ is made small enough, because then the operator $\mathcal{K}_{i}$ becomes
small in comparison to the unit operator $1$. 4. The calculation of the
functions $Y_{i}(r)$ and $Z_{i}(r)$ is not computationally expensive, because
the number of collocation points\ in each partition is prescribed to be small
($16$, usually), and hence the matrices involved, although not sparse, are\ of
small size (e.g. $16\times16$). 5. The accuracy of the calculation of the
functions $Y_{i}(r)$ and $Z_{i}(r)$ can be prescribed ahead of time by
examining the magnitude of the last three coefficients of the expansions. If
they are not smaller than the prescribed accuracy, then the size of the
partition is reduced by a a factor of two, and the accuracy will increase
correspondingly. This adjustment of partition sizes can be done automatically,
as is demonstrated in detail in Ref \cite{AC}.

Next the calculation of the global function $\psi(r)$ in each partition $i$ is
described. Since the functions $Y_{i}(r)$ and $Z_{i}(r)$ are linearly
independent solutions of the Schr\"{o}dinger equation (\ref{SCHR}), and since
the latter is a linear equation, the function $\psi(r)$ can be expressed as a
linear combination of these two functions%
\begin{equation}
\psi(r)=A_{i}Y_{i}(r)+B_{i}Z_{i}(r),~~~b_{i-1}\leq r\leq b_{i}. \label{AB}%
\end{equation}
A relationship between the coefficients $A$ and $B$ in one particular
partition $i$ and those in the other partitions can be obtained by returning
to the original Lippman-Schwinger Eq. (\ref{LS}) for the function $\psi(r),$
with $r$ contained in that particular partition $i$. By expressing the
integrals in Eq. (\ref{LSK}) as sums over the integrals over all partitions,
by inserting for $\psi(r)$ the expression (\ref{AB}) for every partition, and
by making use of Eqs. (\ref{YZ}), one obtains
\begin{equation}
A_{i}=1-\sum_{j\ =\ i+1}^{M}\left[  (GY)_{j}\ A_{j}+(GZ)_{j}\ B_{j}\right]
,~~i=1,2,...M \label{RELAB}%
\end{equation}
and%
\begin{equation}
B_{i}=-\sum_{j\ =\ 1}^{i-1}\left[  (FY)_{j}\ A_{j}+(FZ)_{j}\ B_{j}\right]
,~~i=1,2,...M. \label{RELBA}%
\end{equation}
The $1$ appears in Eq. (\ref{RELAB}) and not in Eq. (\ref{RELBA}) because the
"driving term" in Eq. (\ref{LS}) is $\sin(kr)$ and not $\cos(kr).$ When $i=1$
then the sum in Eq. (\ref{RELBA}) is set to zero, which requires that
$B_{1}=0$. That requirement is compatible with the condition that $\psi(0)=0,$
since $Z_{1}(0)\neq0$ and $Y_{1}(0)=0.$

The equations (\ref{RELAB}) and (\ref{RELBA}) can be manipulated in several
different ways so as to increase the sparseness of the matrices that define
the solutions $A_{i}$ and $B_{i}$. One way, described in Refs. \cite{IEM1} and
\cite{IEM2}, is to subtract from each other Eqs. (\ref{RELAB}) for consecutive
values of $i$, and similarly for Eqs. (\ref{RELBA}). By defining the column
vectors
\begin{equation}
\mathbf{\alpha}_{i}=\left(
\begin{array}
[c]{c}%
A_{i}\\
B_{i}%
\end{array}
\right)  ;~~\mathbf{\omega}=\left(
\begin{array}
[c]{c}%
1\\
0
\end{array}
\right)  ;~~\mathbf{\zeta}=\left(
\begin{array}
[c]{c}%
0\\
0
\end{array}
\right)  \label{VAB}%
\end{equation}
one obtains
\begin{equation}
\left(
\begin{array}
[c]{cccccc}%
\mathbf{I} & \mathbf{M}_{12} &  &  &  & \mathbf{0}\\
\mathbf{M}_{21} & \mathbf{I} & \mathbf{M}_{23} &  &  & \\
& \mathbf{M}_{32} & \mathbf{I} & \mathbf{M}_{34} & \mathbf{..} & \\
&  &  &  &  & \\
&  &  & \mathbf{M}_{M-1,M-2} & \mathbf{I} & \mathbf{M}_{M-1,M}\\
\mathbf{0} &  &  &  & \mathbf{M}_{M,M-1} & \mathbf{I}%
\end{array}
\right)  \left(
\begin{array}
[c]{c}%
\mathbf{\alpha}_{1}\\
\mathbf{\alpha}_{2}\\
\mathbf{\alpha}_{3}\\
..\\
\mathbf{\alpha}_{M-1}\\
\mathbf{\alpha}_{M}%
\end{array}
\right)  =\left(
\begin{array}
[c]{c}%
\mathbf{\zeta}\\
\mathbf{\zeta}\\
\mathbf{\zeta}\\
..\\
\mathbf{\zeta}\\
\mathbf{\omega}%
\end{array}
\right)  \label{M1}%
\end{equation}
where $\mathbf{I}$ and $\mathbf{0}$ are two by two unit and zero matrices,
respectively, and where
\begin{equation}
\mathbf{M}_{i-1,i}=\left(
\begin{array}
[c]{cc}%
(GY)_{i}-1 & (GZ)_{i}\\
0 & 0
\end{array}
\right)  ,~~\ i=2,3,..M \label{M11}%
\end{equation}
and
\begin{equation}
\mathbf{M}_{i,i-1}=\left(
\begin{array}
[c]{cc}%
0 & 0\\
(FY)_{i-1} & (GZ)_{i-1}-1
\end{array}
\right)  ,~~\ i=2,3,..M. \label{M12}%
\end{equation}
Note that Eq. (\ref{M1}) generally connects the $A$ and $B$'s of \emph{three}
contiguous partitions. For example,\textbf{ }$M_{21}\alpha_{1}+\alpha
_{2}+M_{23}\alpha_{3}=\zeta.$

Another way of combining Eqs. (\ref{RELAB} and \ref{RELBA}) is to first write
them into a $(2\times1)$ column form involving the vectors $\mathbf{\alpha
}_{i}$, and subsequently subtracting equations with contiguous $i$-values from
each other, however leaving the last equation in its original form. The result
is \cite{IONEL}
\begin{equation}
\left(
\begin{array}
[c]{cccccc}%
\mathbf{\Gamma}_{1} & \mathbf{-\Omega}_{2} &  &  &  & \\
& \mathbf{\Gamma}_{2} & \mathbf{-\Omega}_{3} &  &  & \\
&  & \mathbf{\Gamma}_{3} & \mathbf{-\Omega}_{4} & \mathbf{..} & \\
&  &  &  &  & \\
&  &  &  & \mathbf{\Gamma}_{M-1} & \mathbf{-\Omega}_{M}\\
\mathbf{\gamma}_{1} & \mathbf{\gamma}_{2} & \mathbf{\gamma}_{3} & .. &
\mathbf{\gamma}_{M-1} & \mathbf{I}%
\end{array}
\right)  \left(
\begin{array}
[c]{c}%
\mathbf{\alpha}_{1}\\
\mathbf{\alpha}_{2}\\
\mathbf{\alpha}_{3}\\
..\\
\mathbf{\alpha}_{M-1}\\
\mathbf{\alpha}_{M}%
\end{array}
\right)  =\left(
\begin{array}
[c]{c}%
\mathbf{\zeta}\\
\mathbf{\zeta}\\
\mathbf{\zeta}\\
..\\
\mathbf{\zeta}\\
\mathbf{\omega}%
\end{array}
\right)  , \label{LDM}%
\end{equation}
where%
\begin{equation}
\mathbf{\Gamma}_{i}=\left(
\begin{array}
[c]{cc}%
1 & 0\\
-(FY)_{i} & 1-(FZ)_{i}%
\end{array}
\right)  , \label{GAMMAi}%
\end{equation}%
\begin{equation}
\mathbf{\Omega}_{i}=\left(
\begin{array}
[c]{cc}%
1-(GY)_{i} & -(GZ)_{i}\\
0 & 1
\end{array}
\right)  , \label{OMEGAi}%
\end{equation}
and%
\begin{equation}
\mathbf{\gamma}_{i}=\left(
\begin{array}
[c]{cc}%
0 & 0\\
(FY)_{i} & (FZ)_{i}%
\end{array}
\right)  . \label{gammai}%
\end{equation}
It is noteworthy that the first $M-1$ equations in (\ref{LDM}),%
\begin{equation}
\mathbf{\ }\Gamma_{i}\ \alpha_{i}=\Omega_{i+1}\alpha_{i+1},~~~i=1,2,..M-1
\label{MATCH}%
\end{equation}
are equivalent to matching the wave function $\psi$ at the end of partition
$i$ to $\psi$ at the start of partition $i+1.$ This can be seen by imposing
the two conditions $\psi_{i}(b_{i})=$ $\psi_{i+1}(b_{i})$ and $\psi
_{i}^{\prime}(b_{i})=$ $\psi_{i+1}^{\prime}(b_{i})$ where $\psi_{i}$ is the
wave function in partition $i$ given by Eq. (\ref{AB}) and where $\psi
_{i}^{\prime}$ is the corresponding derivative. Inserting into Eq. (\ref{AB})
the values of $Y_{i}$ and $Z_{i}$ or their derivatives at either the beginning
or the end of a partition as given by Eqs. (\ref{Bi-1}) or (\ref{Bi}),
respectively, one obtains the result
\begin{align*}
A_{i}  &  =A_{i+1}[1-(GY)_{i+1}]-B_{i+1}(GZ)_{i+1}\\
B_{i+1}  &  =-A_{i}(FY)_{i}+B_{i}[1-(FZ)_{i}].
\end{align*}
These two equations are equivalent to Eq. (\ref{MATCH}) .

By successive applications of Eq. (\ref{MATCH})
\[
\mathbf{\alpha}_{i+1}\mathbf{=}\left(  \Omega_{i+1}\right)  ^{-1}%
\mathbf{\ \Gamma}_{i}\mathbf{\ \alpha}_{i}%
\]
one can relate the values of $\alpha_{i},$ $i=2,3,..M,$ to $\alpha_{1}$ and
then use the last of the (\ref{LDM}) equations
\begin{equation}
\sum_{i=1}^{M-1}\ \mathbf{\gamma}_{i}\mathbf{\ \alpha}_{i}\mathbf{+\alpha}%
_{M}=\left(
\begin{array}
[c]{c}%
1\\
0
\end{array}
\right)  \label{SUM}%
\end{equation}
in order to find the value of $A_{1}.$ It can be shown that Eq. (\ref{SUM}) is
compatible with the requirement that $B_{1}=0.$

Several comments are in order.\newline a) The "big" matrices in Eqs.
(\ref{LDM}) or (\ref{M1}) are sparse, and can be solved by Gaussian
elimination. Since the number of floating point operations (flops) is of order
$M$, the computational complexity of the S-IEM is comparable to that of the
solution of the differential equation. This sparseness property results from
the semi-separable nature of the integration kernel $\mathcal{K},$ as is shown
in Refs. \cite{IEM1}, \cite{IEM2}, which however applies only in the
configuration representation of the Green's function. This part of our
procedure also differs substantially from that of Ref. \cite{FR}. \newline b)
The scattering boundary conditions can be implemented reliably. This is
because the Greens function incorporates the asymptotic boundary conditions
automatically. However, in the coupled channel case for angular momentum
numbers $L>0,$\ the coupled equations have to be solved as many times as there
are open channels because our Green's functions are composed of $\sin
(kr)$\ and $\cos(kr),$ rather than of Riccati-Bessel functions. We show
\cite{IEM2} that the desired linear combination of the solutions can be
obtained without appreciable loss of accuracy, since the matrix required in
the solution for the coefficients has a condition number not much larger than
unity. This means that our various solutions are linearly independent to a
high degree, contrary to what can be the case with the solution of
differential equations. \newline c) The method is very economical in the total
number of mesh-points required in the interval $[0,T]$ because in each
partition or spectral collocation method requires very few mesh points (like
in the case of Gauss-Legendre integration as compared to Simpson'
integration), and the required length of each partition can be easily adjusted
to optimal size based on the magnitude of the coefficients of the expansion of
the functions $Y$ and $Z$ into Chebyshev polynomials, as described
before.\newline d) The calculation can be distributed onto parallel
processors. This is because the functions $Y$ and $Z$, as well as the overlap
integrals (\ref{OYZ}), required for Eqs. (\ref{LDM}) or (\ref{M1}), can be
calculated separately for each partition independently of the other ones. This
is an important point, since if the number of channels increases, the number
of the quantities (\ref{OYZ}) increases accordingly.

Property c) is also important because, due to the small number of total
mesh-points, the accumulation of machine round-off errors is correspondingly
small. In addition, as is well known, integration is numerically more stable
than differentiation as discussed for example in sections 4.4 and 5.2 on pages
203 and 263, respectively, in Ref. \cite{BF}, and is also shown in Tables
\ref{TAB2} and \ref{TAB4}. Hence the accumulation of the inherent round-off
error is smaller for the numerical solution of an integral equation than for
the numerical solution of differential equations. The small accumulation of
roundoff errors in comparison to a finite difference method is clearly
illustrated in Fig. 1 of Ref. \cite{IEM1}, which compares the round off errors
in the solution of Bessel's equation obtained via the IEM with that of the
Numerov method.

\section{Applications}

\begin{figure}
[ptb]
\begin{center}
\includegraphics[
natheight=3.392300in,
natwidth=3.661300in,
height=3.5442in,
width=3.8232in
]%
{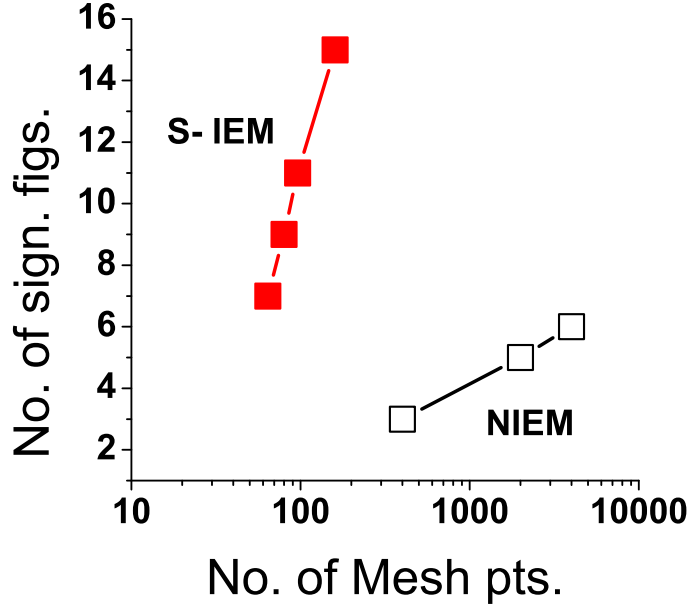}%
\caption{Comparison of the numerical stability of two methods for calculating
the singlet phase shift for electron-hydrogen scattering, as described in the
text. The number of significant figures on the y-axis is the number of decimal
places for which the result remains the same as the number of meshpoints is
increased. S-IEM is the the spectral method described in this paper, and NIEM
is a non-iterative method of solving the same integral equation carried out by
Sams and Kouri.}%
\label{FIG1}%
\end{center}
\end{figure}

The various features of the S-IEM method will now be illustrated by means of
examples. The spectral property that high accuracy is reached very rapidly (in
principle faster than any inverse power of the number of mesh-point in a given
radial interval) is illustrated for the case of the scattering of an electron
from an Hydrogen atom. This is a suitable example, because the identity
between the incoming electron and the electron bound in the atom leads to an
additional integral term in the Schr\"{o}dinger equation, if the Pauli
exclusion principle is implemented via the Hartree-Fock formulation.
Rigorously including this term is difficult for the conventional finite
difference methods, and various techniques were developed for that purpose
\cite{NIEM}, and additional references can be found in \cite{EXCH}. By
contrast, in the IEM method this additional integral term is easily
incorporated without substantial loss of accuracy \cite{EXCH}, because the
integral kernel is semi-separable. A comparison between the S-IEM and a
conventional NIEM method \cite{kouri} is shown in Fig (\ref{FIG1}).%
The $L=0$ singlet phase shift was calculated for the incident momentum
$\ k=0.2\ (a_{0})^{-1}$ and $T=50\ a_{0}$, while the target electron was kept
in the ground state of the Hydrogen atom. The figure shows that, as the number
$m$ of partitions is increased, and accordingly the number of mesh-points
$m\times16,$\ the number of stable significant figures in the phase shift
increases very rapidly for the S-IEM, illustrating the spectral nature of that
method. By comparison, for a method employing finite difference techniques
based on an equi-spaced set of mesh-points, the number of stable significant
figures increases much more slowly \cite{kouri} for solving a very similar
integral equation non-iteratively by means of the NIEM method. Although it
gives a good illustration of the numerical accuracy, this example is
nevertheless not very realistic physically because the virtual excitations of
the bound electron to the myriad of possible states, both bound and in the
continuum, is not included. Inclusion of these excitations requires "state of
the art" calculations that are presently in progress \cite{ART}.%

\begin{figure}
[ptb]
\begin{center}
\includegraphics[
natheight=3.286800in,
natwidth=3.863800in,
height=3.4338in,
width=4.0324in
]%
{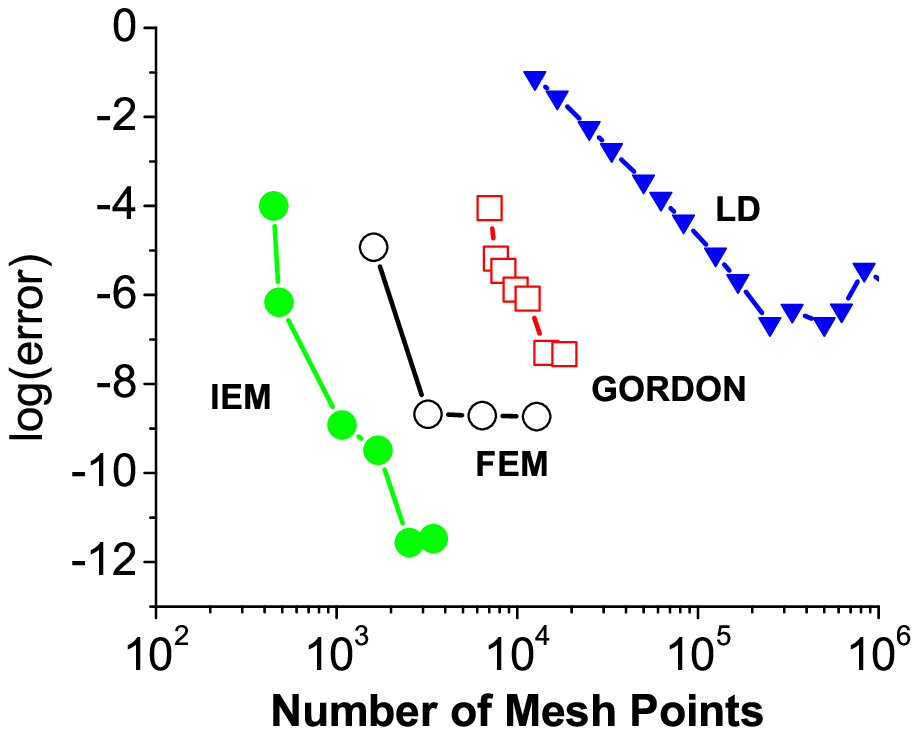}%
\caption{Comparison of errors for various methods of computation of the $L=0$
phase shift for cold atom collision, as a function of the number of mesh
points in a fixed radial interval. IEM is the method described here, FEM is a
finite element method, Gordon and LD (logarithmic derivative) are two finite
difference methods, as explained in the text.}%
\label{FIG2}%
\end{center}
\end{figure}

Another example is the scattering of atoms at ultra-low temperature. This
information is needed for the investigation of photo association \cite{PHOTOA}
of the two atoms into a molecule, and also in the formation of Bose-Einstein
condensates (BE) \cite{BE}. The lifetime of a BE condensate is
reduced\ \cite{BE2} by the three-body process in which two of the atoms
combine to form a molecule in the presence of a third atom, that in turn
carries away the energy of formation of the dimer. The depletion rate is
proportional to the fourth power of the scattering length. At low energies a
stable method of calculation is required because, the lower the incident
energy, the more the long-range part of the potentials contributes
significantly to the phase shift. A bench mark calculation was performed using
the S-IEM method, involving two channels, one closed and one open \cite{AC}.
\ The numerical stability of the $L=0$ scattering phase shift$\ $as a function
of the number of mesh points used was investigated, and was compared with
various other methods of calculation, and the results are shown in Fig.
\ref{FIG2}. In all of these calculations the maximum radius is $T=500$ atomic
units ($a_{0}$ or\ $Bohr),$ the diagonal potentials are of the Lenard Jones
form $C_{6}/r^{6}+C_{12}/r^{12},$ and the coupling between the two channels is
of an exponential form \cite{AC}. At small distances, due to the large depth
of the potentials, the wave function oscillates rapidly, and hence it is
important to be able to adjust the size of the partitions accordingly. Since
no analytical exact comparison values exist, the "error" in the figure is
defined as the absolute value of the difference between the result for a given
value of the number of mesh points $N$ and the maximum value of $N$ employed
in the particular method. The FEM method is a finite element method \cite{FE}
implemented by B. D. Esry and carried out by J. P. Burke , Jr \cite{AC}; the
Gordon method \cite{GORD} was implemented by F. E. Mies \cite{AC}, and LD is a
logarithmic derivative method implemented by the code MOLSCAT \cite{MOL},
\cite{IS}. For the \ LD curve the roundoff errors apparently overwhelm the
truncation errors when the number of mesh points is larger than $2\times
10^{5}$. The S-IEM again shows a rapid improvement of accuracy with the number
of mesh-points, and it reaches a somewhat higher stability than the FEM. Our
bench mark calculation was recently used \cite{IXARU} for comparison with a
finite difference method in which the potential in each partition is assumed
constant (similar to what is the case with one form of the Gordon method), and
the corrections are taken into account iteratively.%

\begin{figure}
[ptb]
\begin{center}
\includegraphics[
natheight=3.272700in,
natwidth=3.782500in,
height=3.4188in,
width=3.9477in
]%
{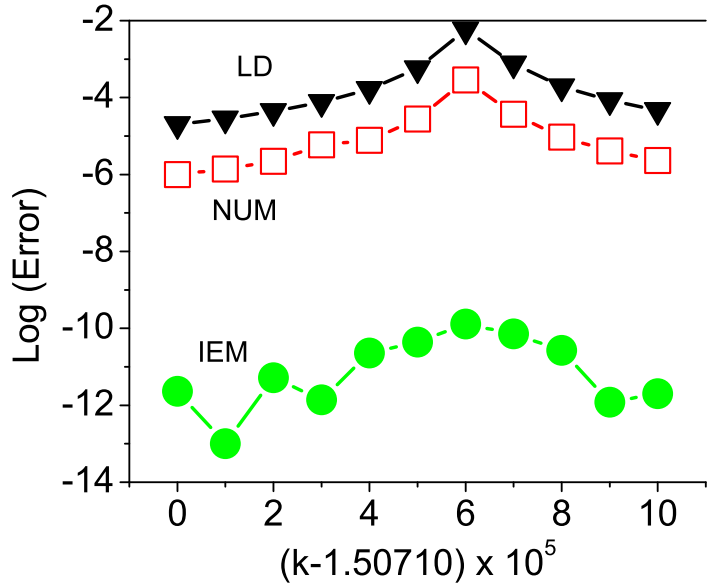}%
\caption{Numerical error in the phase shift for scattering from a Morse
potential with a barrier, in the region of a narrow resonance, as a function
of the incident momentum. The error is obtained by comparison with the
analytic result, the momentum closest to the resonance occurs for
$k=1.50716.$}%
\label{FIG3}%
\end{center}
\end{figure}

In many quantum mechanical calculations, penetration of the wave function
through a barrier is involved. Examples in nuclear physics are the alpha
particle decay of a nucleus, or the fission of a nucleus into two daughter
nuclei, or in the scattering of a nucleus by another nucleus, and also in many
similar situations in atomic physics. A barrier frequently occurs when a long
range repulsive potential, such as a centripetal potential of the form
$L(L+1)/r^{2}$ or that of a repulsive Coulomb potential, is added to an
attractive nuclear or atomic potential of a shorter range. For the scattering
or the fusion reaction of a light nucleus with a heavy nucleus at low incident
energies \cite{KOLA}, \cite{MICHEL} the penetration of the corresponding wave
function through such a barrier can pose substantial calculational challenges
\cite{MAGDA}. In low temperature atom-molecule scattering, similar barrier
penetration effects become crucial \cite{DALGARNO}. For this reason a test of
the accuracy of a calculation for a case involving barrier penetration was
performed. The potential chosen is an "inverted" form of the Morse potential
\cite{M} for which analytic results exist for the scattering phase shift
\cite{MORSE}. It has an attractive negative valley near the origin at $r=0$
followed by a smooth positive energy barrier, a situation which leads to
resonances. For resonant energies the wave function in the valley region can
become very large if the width of the resonance is sufficiently small, and in
the barrier region this wave function \emph{decreases} as a function of
distance. This decrease of the wave function in the barrier region amplifies
the numerical errors, since in this region the numerical errors tend to
\emph{increase} exponentially. The accuracy of three methods of calculation
for a\ particular resonance which occurs for an incident momentum $k$ in the
region $1.5071fm^{-1}<k<1.5072fm^{-1}$ are illustrated in Fig. \ref{FIG3}. The
parameters of the Morse potential are given in Fig.$10$ of Ref. \cite{MORSE},
the maximum amplitude of the wave function in the valley region at the
resonance near $k=1.50716fm,$ is close to $300$ (asymptotically it is equal to
$1$). The error is defined as the difference between the analytical and the
numerical results; the momenta $k$ on the x-axis are given as the excess over
the momentum at the left side of the resonance, $k=1.50710\ fm^{-1}$. The IEM
curve is obtained with the method described in this paper, NUM is a sixth
order Numerov method, also denoted as Milne's method \cite{EZ}, and the LD
curve is obtained with the Logarithmic Derivative method, implemented by
MOLSCAT \cite{B}. The matching radius for the two finite difference methods,
LD and NUM, was set at $50\ fm,$ and the corresponding analytical values were
extrapolated from $T=\infty$ to $T=50\ fm$ by a Green's function iteration
procedure described in Ref. \cite{AC}, and are listed in Table 1 of Ref.
\cite{MORSE}. For the more precise S-IEM calculation that extrapolation was
not accurate enough, and $T=100$ was used instead. One sees from the figure
that the accuracy of the S-IEM is several order of magnitudes (six) higher
than that of the Numerov method.

\section{Discussion and Conclusions}

A simple way to distinguish a spectral method from a finite difference method
is that, in a particular partition, the mesh points in the former are not
equi-spaced, while in the latter they are. Even though the accuracy of finite
difference methods can be substantially increased by extrapolating the
algorithms to equivalent zero-sized distance between mesh points \cite{GRAGG},
such extrapolation methods may become cumbersome. Our spectral S-IEM method is
one of a class of well-known methods that divide the spatial domain into
partitions (or sectors), and expand the solution on a suitable set of basis
functions in each partition. One example is the method of Gordon \cite{GORD},
that uses Airy basis functions. The potential in each partition is
approximated by a linear function, and the Airy functions are the
corresponding exact solutions of the differential equation. This method was
included among the comparisons carried out for the atom-atom scattering case,
illustrated in Fig. \ref{FIG2}. Gordon's method is widely used for atomic
physics calculations, and one of the implementations can be found in Refs.
\cite{THOMPSON} and \cite{IXARU}. This is a "potential following method" that
is particularly efficient when the potential varies slowly with distance.
Another example is the method utilized by Light and Walker \cite{LW} in which
the potential in each partition is approximated by a constant. In this case
the Green's function that propagates the solution from one end of the
partition to the other can be written simply in terms of sine and cosine
functions. This method lends itself well to propagate the inverse of the
logarithmic derivative of the solution from one end of a partition to the
other end, without calculating the solution itself. This is called the
R-matrix propagation method, and has been implemented by Burke and Noble
\cite{BN}. This method, as implemented by the code MOLSCAT \cite{MOL}, was
included among the comparisons carried out for the barrier penetration
calculation, illustrated in Fig. \ref{FIG3}. A "function following" method
that expands the Greens function in a given partition in terms of Legendre
Polynomials, without making approximations on the potentials, is given by
Baluja \emph{et al.} \cite{BBM}. This method is also implemented in the
computer code FARM \cite{BN}. The resulting expansion of the distorted Green's
function $\mathcal{G}(r,r^{\prime})$ is of a separable form, i.e., it is given
as a sum over products of functions $u(r)\times v(r^{\prime}).$A similar form
is obtained by using Sturmian basis functions \cite{STURMIANS}, \cite{CANTON}.
However such expansions do not converge to high accuracy because the
derivative of a Green's function has a discontinuity at the points
$r=r^{\prime}$. Our S-IEM method does not suffer from that difficulty because
the distorted Green's function $\mathcal{G}(r,r^{\prime})$ is obtained in
terms of the exact undistorted Green's function $\mathcal{G}_{0}(r,r^{\prime
})$ through Eq. (\ref{YZ}). The numerical solution of Eq. (\ref{YZ}) is
equivalent to expressing the distorted Green's function in terms of the
undistorted one, according to%
\[
\mathcal{G=(}1-\mathcal{G}_{0}V)^{-1}\mathcal{G}_{0},
\]
and since\ $\mathcal{G}_{0}$ is given exactly in terms of its semi-separable
form [near Eq. (\ref{LS})] there is no loss of accuracy. The functions $Y(r)$
and $Z(r)$ are two independent solutions of both the Schr\"{o}dinger equation
and the Lippman-Schwinger equation in a particular partition, and they
represent the two basis functions in terms of which the global solution is
obtained in each partition. The equation (\ref{MATCH}), based on algebraic
matrix Eq. (\ref{LDM}), that relates the two expansion coefficients in one
partition to the coefficients of \emph{one} adjoining partition is equivalent
to the propagation of the logarithmic derivative from one partition to the
next. However, the method represented by Eq. (\ref{M1}) relates the
coefficients in one partition to those in \emph{two} other partitions appears
not to be as closely related to the propagation of the logarithmic derivative,
hence a comparison of the two methods for particular cases would be very
desirable. The method involving two adjoining partitions can be shown to be
very similar to the multiple shooting method for solving two-point boundary
vaue problems \cite{SB}.\ How the computational complexity scales with the
number of coupled channels, in comparison with that of other methods, has also
yet to be investigated.

In summary, a recently developed method for solving the Lippman-Schwinger
integral equation is described and is applied to the solution of several
physical problems. Since the new S-IEM is considerably more stable than finite
difference methods, it is concluded that the S-IEM may become the method of
choice for particular applications, such as atomic physics calculations that
involving large distances, require high accuracy, and need to be carried out
in configuration space.


\begin{thebibliography}{99}
\bibitem {IS}The authors are thankful to Dr. Ionel Simbotin, at the Dept. of
Physics, University of Connecticut, for having performed the calculation with
the MOLSCAT code.
\end{thebibliography}
\end{document}